\documentclass[article]{aa} 
\usepackage{graphicx}
\usepackage{txfonts}

\newcommand\simlt{\lower.5ex\hbox{$\; \buildrel < \over \sim \;$}}
\newcommand\simgt{\lower.5ex\hbox{$\; \buildrel > \over \sim \;$}}

\newcommand\be{\begin{equation}}
\newcommand\ee{\end{equation}}

\newcommand{\Bbeta}{\mbox{\boldmath$\beta$}}

\begin{document}

\title{Polarization in the inner region of Pulsar Wind Nebulae}

\author{
N. Bucciantini\inst{1}, L. Del Zanna\inst{2}, E. Amato\inst{3}, D. Volpi\inst{2}
}
\institute{
Astronomy Department, University of California at Berkeley, 601Campbell Hall, 94720 Berkeley, CA, US \\ \email{niccolo@astron.berkeley.edu}
\and 
Dipartimento di Astronomia e Scienza dello Spazio, Universit\`a di Firenze, Largo E.Fermi 2, I-50125 Firenze, Italy
\and 
INAF, Osservatorio Astrofisico di Arcetri, Largo E.Fermi 5, I-50125 Firenze, Italy
}

\date{Received 20 June 2005/ Accepted 27 July 2005 }

\abstract{We present here the first effort to compute synthetic synchrotron polarization maps of Pulsar Wind Nebulae (PWNe). Our goal is to highlight how polarization can be used as an additional diagnostic tool for the flow structure in the inner regions of these nebulae. Recent numerical simulations suggest the presence of flow velocities $\sim 0.5\; c$ in the surroundings of the termination shock, where most of the high energy emission comes from. We construct polarization maps taking into account relativistic effects like Doppler boosting and position angle swing. The effect of different bulk velocities is clarified with the help of a toy-model consisting of a uniformly emitting torus. We also present a map based on recent numerical simulations of the entire nebula and compare it with presently available data. The comparison with upcoming high resolution observations could provide new insight into the inner structure of the nebula and put constraints on the geometrical properties of the magnetic field.
\keywords{Stars: pulsar: general - ISM: supernova remnant - Radiation mechanism: non-thermal - Polarization - magnetohydrodynamics (MHD) - relativity}
}

\maketitle

\section{Introduction}

Pulsar Wind Nebulae (PWNe) are bubbles of relativistic particles and magnetic field produced by the interaction of the relativistic pulsar wind with the ambient medium (ISM or Supernova Remnant). These bubbles are souces of non thermal (synchrotron and Inverse Compton) emission, in a broad range of frequencies from radio wavelengths to gamma rays. X-ray images of PWNe show an axisymmetric feature known as the {\it jet-torus} structure. This feature have been observed by now in a number of PWNe, among which the Crab Nebula (Weisskopf et al. \cite{weiss00}), Vela (Helfand et al. \cite{helfand01}) and PSR B1509-58 (Gaensler et al. \cite{gaensler02}). Recent relativistic MHD simulations (Komissarov \& Lyubarsky \cite{kom03}, Komissarov \& Lyubarsky \cite{kom04}, Del Zanna et al. \cite{luca04}) have shown that this structure naturally arises if the pulsar wind energy flux is anisotropic, higher in the equatorial plane of the pulsar rotation than along the polar axis, and the ratio of Poynting to particle energy flux, $\sigma$, is greater than approximately $1\%$. We will not discuss here the details of those simulations and the related flow structure inside PWNe. Let us just recall that the derived synthetic synchrotron maps allow to recover, at least qualitatively, the torus structure with a bright {\it main arc} and other features, like the {\em knot} that is observed in the Crab Nebula almost coincident with the pulsar position.

Despite the success of these models at reproducing most of the basic observational features, there are still details that cannot be accounted for because of the intrinsic model assumptions. A major limitation appears to be the hypothesis of a purely toroidal field, which vanishes on the axis, thus strongly suppressing the jet emissivity. In order to reproduce the observed PWN jets an additional magnetic field component, either poloidal or disordered, must be necessarily introduced. At present, in the absence of fully 2.5-D, or even 3-D, MHD simulations, only {\em ad-hoc} prescriptions have been used. Here we look for constraints that can be put on the geometry of the magnetic field, in particular the presence and strength of a poloidal component.

As shown by Shibata et al. \cite{shibata03}, a completely disordered (isotropic) magnetic field has a very different emission signature with respect to a purely toroidal one, in the case of low bulk speed in the torus. However, the flow speed in the torus region, both from observations and the above cited MHD simulations, is of the order of $\sim 0.5\, c$ and Doppler boosting effects strongly affect the overall emission, making it difficult to constrain the field structure.

On the other hand, polarization is supposed to provide a more direct way to investigate the magnetic field configuration. Radio polarization maps are available for many PWNe, among which the Crab Nebula (Wilson \cite{wilson72}, Velusamy \cite{velusamy85}), 3C58 (Wilson \& Weiler \cite{wilson76}, Weiler \& Seiesland \cite{weiler71}), G328.4+0.2 (Gaensler et al. \cite{gaensler00}), Vela (Dodson et al. \cite{dodson03}), SNR G320.4-01.2 (Gaensler et al. \cite{gaensler99}), but radio emitting particles have too long synchrotron cooling times, so that the nebular emission is usually dominated by the outer regions (the Crab Nebula torus is barely visible in radio). Radio polarization might then just provide a good estimate of the degree of ordered versus disordered magnetic field for the overall nebula, but cannot be used to constrain the region close to the termination shock. What one would need is a polarization map at higher frequencies. In the X-rays, where the jet-torus structure was first clearly recognized, polarimetry is unfortunately not available at present. However, the torus and other bright features near the termination shock (like the {\it inner ring}, the time-dependent {\it wisps}, and the central {\it knot}) are well visible in the Crab Nebula at optical wavelengths, at which high resolution polarimetry might be employed. Until now only few optical polarization maps of the Crab Nebula (Schmidt et al. \cite{schmidt79}, Hickson \& van der Bergh \cite{hickson90}, Michel et al. \cite{michel91}) have have been published, in general with too low spatial resolution to investigate the inner region in detail, though recent {\it Hubble} data appear to be more promising (James Graham, personal communication).

Here we present synthetic polarization maps as a novel diagnostic tool for the physics of PWNe. We will first discuss the basic ingredients that determine the polarization direction and fraction within the framework of a simplified toy-model. Then this tool will be applied to the flow properties derived from (axisymmetric) relativistic MHD simulations, as those presented in Del Zanna et al. \cite{luca04}. Our work is aimed at clarifying if and how polarization maps might be used to infer information on the inner flow structure, by comparison with high-resolution optical observations.

In Sect.~2 we describe the model adopted for the emission and for the polarization, both for the toy-model and in the case of the more complex flow pattern derived from numerical simulations. The results are discussed and the effect of relativistic bulk flow speeds on the polarization are shown. Finally in Sect.~3 we summarize our conclusions.

\section{Synchrotron emissivity and polarization}
Let us briefly recall here the derivation of the local synchrotron emissivity $s(\nu)$, at a given frequency $\nu$, for the case in which the particles' energy distribution function at the termination shock is a power law
\be
f(\epsilon_0)\propto P_0\,\epsilon_0^{-(2\alpha+1)},
\ee
where $P_0$ is the total thermal pressure immediately downstream of the shock. As shown in Bucciantini et al. \cite{me05} (to which the reader is referred for a detailed derivation of the following equations), for $\alpha$ close to 0.5 the emissivity at a generic position inside the nebula becomes
\be
s(\nu)\propto P D^{\alpha+2}(B'_\perp)^{\alpha+1} \nu^{-\alpha}
\label{eq:snu}
\ee
where $P$ is the local thermal pressure, $B'_\perp$ the magnetic field component perpendicular to the line of sight as measured in the fluid frame, and $D$ is the Doppler boosting factor.

Since we are interested to high energy emitting particles, whose characteristic lifetime is comparable to or shorter than the age of the nebula, synchrotron losses must be taken into account. Hence we correct the emissivity by multiplying the Eq.~\ref{eq:snu} by:
\be
[1-\epsilon (\nu)/\epsilon_\infty]^{2\alpha-1},
\label{losses}
\ee
where $\epsilon (\nu)$ is the energy of particles emitting at frequency $\nu$ (averaged over the pitch angle for an isotropic distribution) and $\epsilon_\infty$ is the local energy of a particle injected at the termination shock with infinite energy (we assume zero emissivity where $\epsilon_\infty \leq\epsilon(\nu)$). The nebular emission map is then finally constructed by integrating along the line of sight.

In order to build a polarization map, given a pattern of velocity streamlines, one also needs to evaluate the Stokes parameters $U, Q, V$. However, since synchrotron radiation from high Lorentz factor particles is known to have a very low degree of circular polarization, here we will neglect it completely ($V=0$) and thus evaluate only $U$ and $Q$.

By naming ${\rm d}I$ the total intensity of the radiation emitted by a fluid element (the emissivity $s(\nu)$ in Eq.~(\ref{eq:snu})), the value of ${\rm d}Q$ and ${\rm d}U$ are defined in terms of ${\rm d}I$ and the polarization angle $\chi$ as:
\begin{eqnarray}
{\rm d}Q=\frac{\alpha+1}{\alpha+5/3}{\rm d}I\cos{2\chi},\\
{\rm d}U=\frac{\alpha+1}{\alpha+5/3}{\rm d}I\sin{2\chi}.
\end{eqnarray}
We define $\chi$ as the angle, in the plane of the sky, between the emitted electric field and the projected axis of the nebula. Its value is computed in two steps, as in Pariev et al. \cite{pariev03} (see also Blandford \& K\"onigl \cite{blandford79}, Bj\"ornsson \cite{bjornsson82}, Qian \& Zhang \cite{qian03}). First we evaluate the inclination angle $\psi$ of the emitted radiation electric field ${\bf e}$ with respect to a plane containing the observer direction ${\bf{n}}$ and the velocity vector $\Bbeta$ (considering that in the comoving frame ${\bf e}$ is normal both to the local magnetic field and to ${\bf{n}}$). By choosing a Cartesian coordinate system with the $x$ axis in the direction of $\Bbeta$ and $y$ parallel to $\bf{n}\times\Bbeta$ we find
\be
\tan{\psi}=\frac{B_z}{B_y}\frac{|\Bbeta|-\Bbeta\cdot {\bf{n}}/|\Bbeta|}
{1-\Bbeta\cdot {\bf{n}}},
\label{eq:psi}
\ee
where we have assumed $\Bbeta\cdot{\bf{B}}=0\Rightarrow B_x=0$, since our simplified model and the MHD simulations both assume a geometry with a purely toroidal field and a purely poloidal velocity. Note that the above equation differs from Eq. (24) in Qian \& Zhang \cite{qian03}, because of a mistake in the definition of the magnetic field components. Finally $\chi$ is obtained by adding to $\psi$ a quantity representing the angle between the plane containing $\Bbeta$ and ${\bf{n}}$, and the plane containing ${\bf{n}}$ and the projected axis of the nebula. Once the value of $\chi$ is known for the radiation emitted by a fluid element, the Stokes parameters are computed by integrating the equations for ${\rm d}Q$ and ${\rm d}U$ along the line of sight, thus providing maps of both polarization fraction and position angle.

\subsection{Uniform torus model}
For the sake of clarity let us first consider the simple case where the emission is concentrated in a homogeneously radiating torus around the pulsar. The existence of arcs and ring-like features observed in a number of PWNe may be modeled as a first approximation in this way. In a spherical reference system ($r,\theta,\phi$) centered on the pulsar position the torus can be shaped as:
\be
(r\sin{\theta}-R_1)^2+(r\cos{\theta})^2-R_2^2\le 0,
\ee
where $R_1$ is the torus principal radius and $R_2$ is the radius of the cross section (both will be given here in arbitrary units). Inside this region we consider quantities with homogeneous values, with the magnetic field assumed to be purely toroidal, whereas the flow velocity is assumed purely radial. In this simplified model we do not take into account the effect of synchrotron losses, that is the correction in Eq.~(\ref{losses}), thus the resulting maps are for a generic frequency. Outside the torus we impose zero emissivity. 

\begin{figure*}
\centering
    \resizebox{\hsize}{!}{\includegraphics{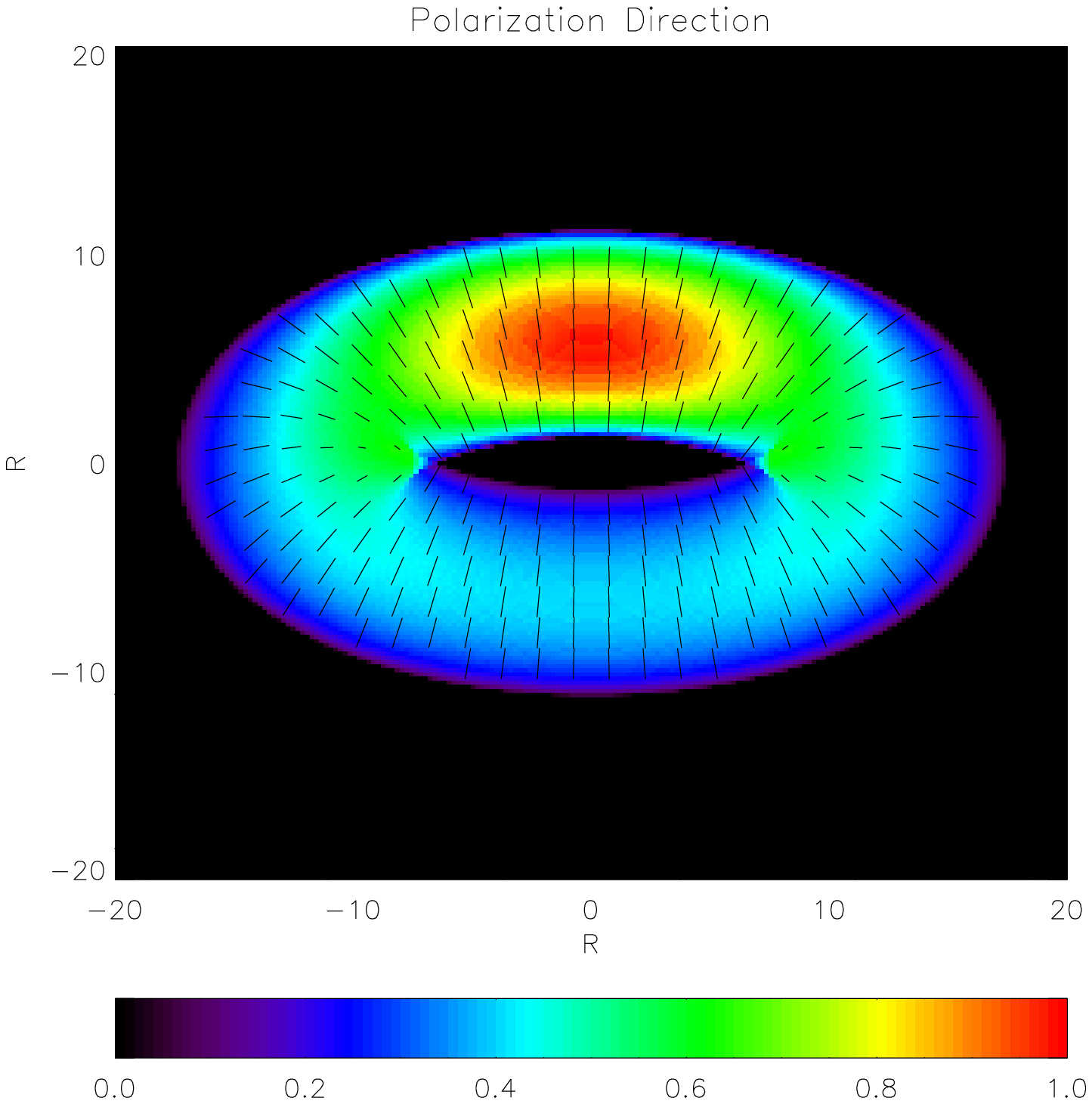}\includegraphics{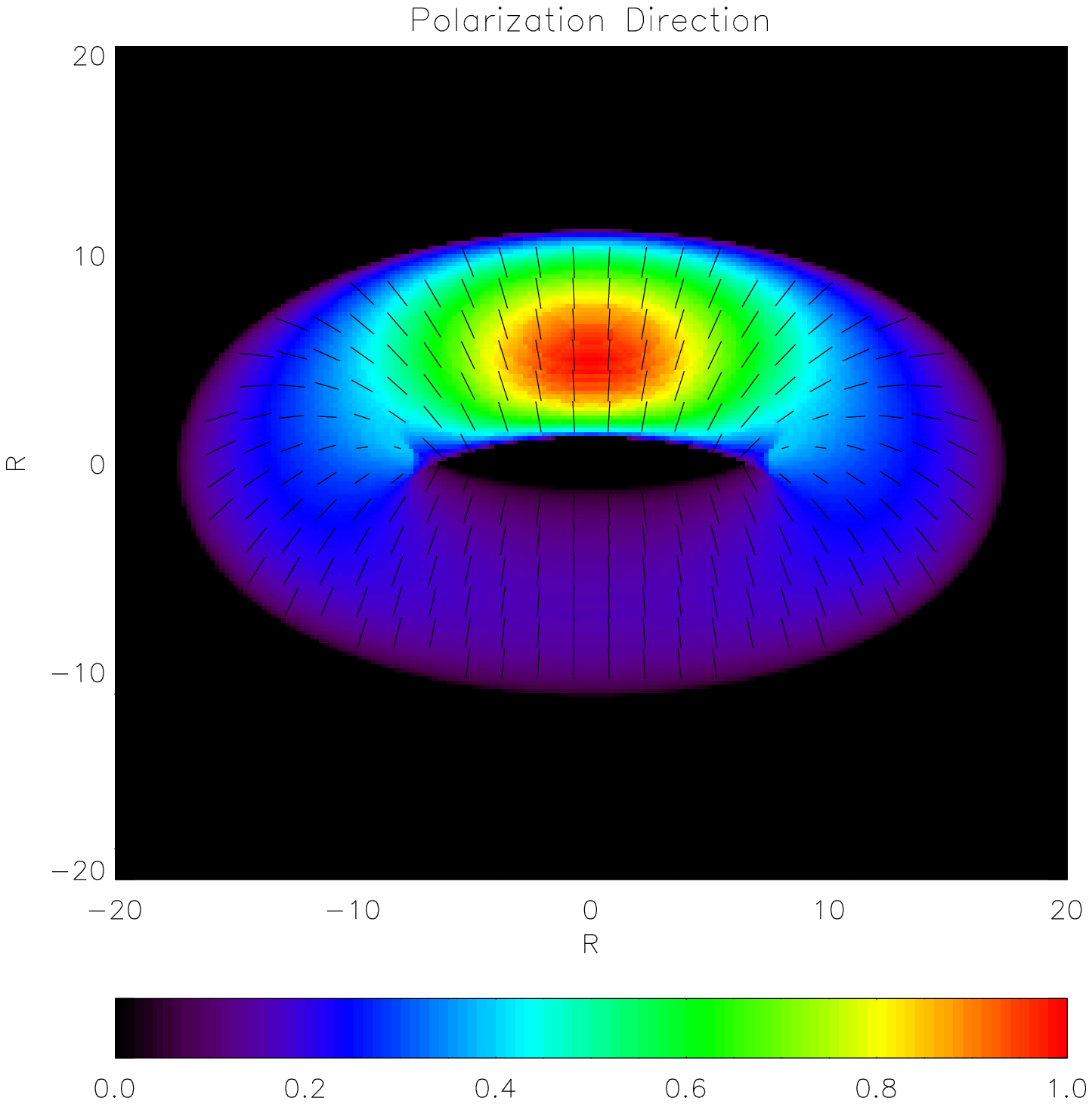}}
    \resizebox{\hsize}{!}{\includegraphics{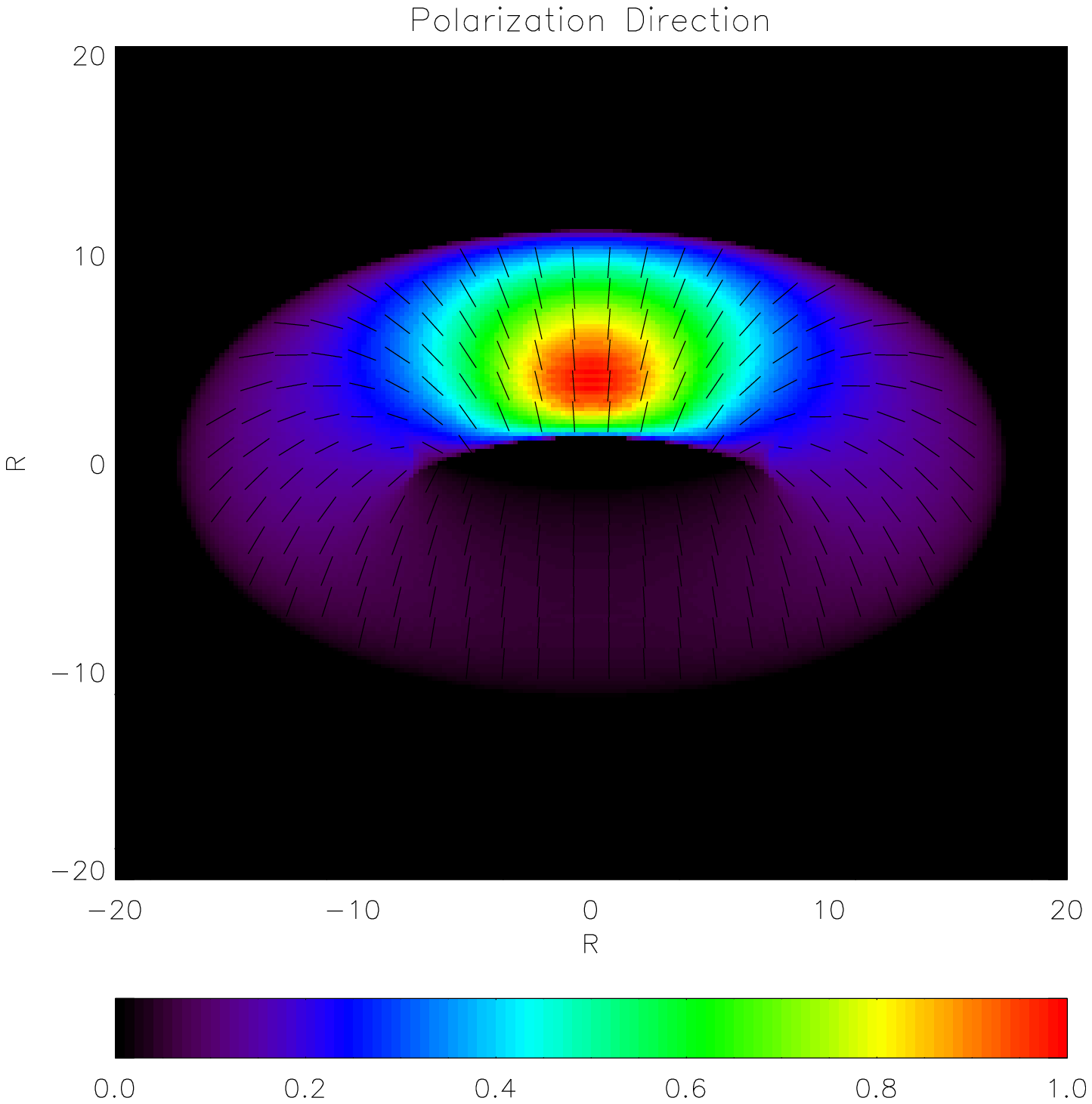}\includegraphics{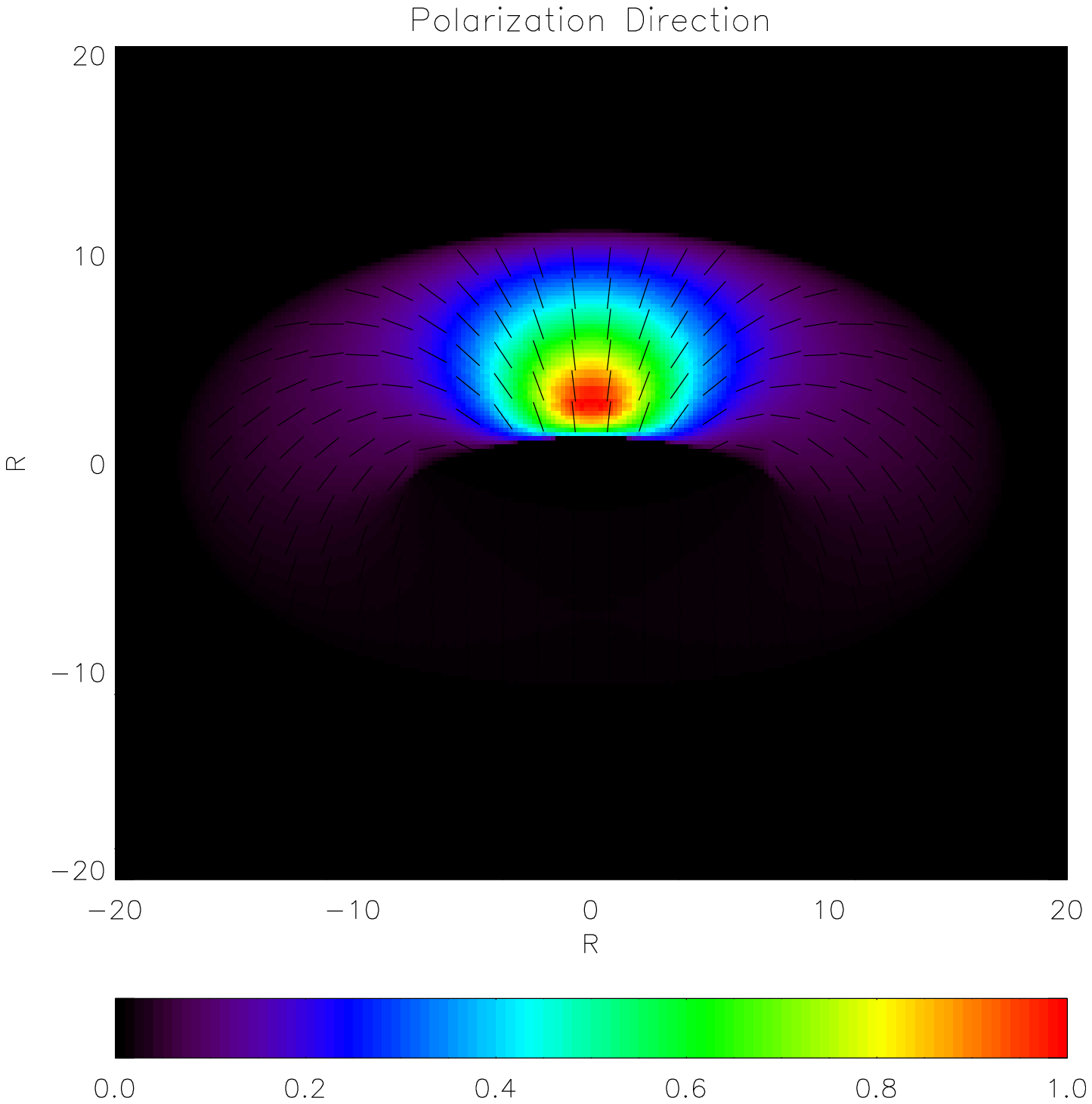}}
   \caption{Polarization vector and total emission (normalized to the maximum) in the uniform torus case (here $R_1=12.5$ and  $R_2=5$, in arbitrary units, chosen to reproduce approximately the ratio $R_2/R_1$ estimated for the Crab Nebula torus). The inclination angle of the symmetry axis with respect to the plane of the sky is $\delta=30^\circ$. Top-left $v=0.2\, c$; top-right $v=0.4\, c$; bottom-left $v=0.6\, c$; bottom-right $v=0.8\, c$. The length of the polarization vector is proportional to the polarization fraction.}
   \label{fig:1}
\end{figure*}

\begin{figure*}
\centering
    \resizebox{\hsize}{!}{\includegraphics{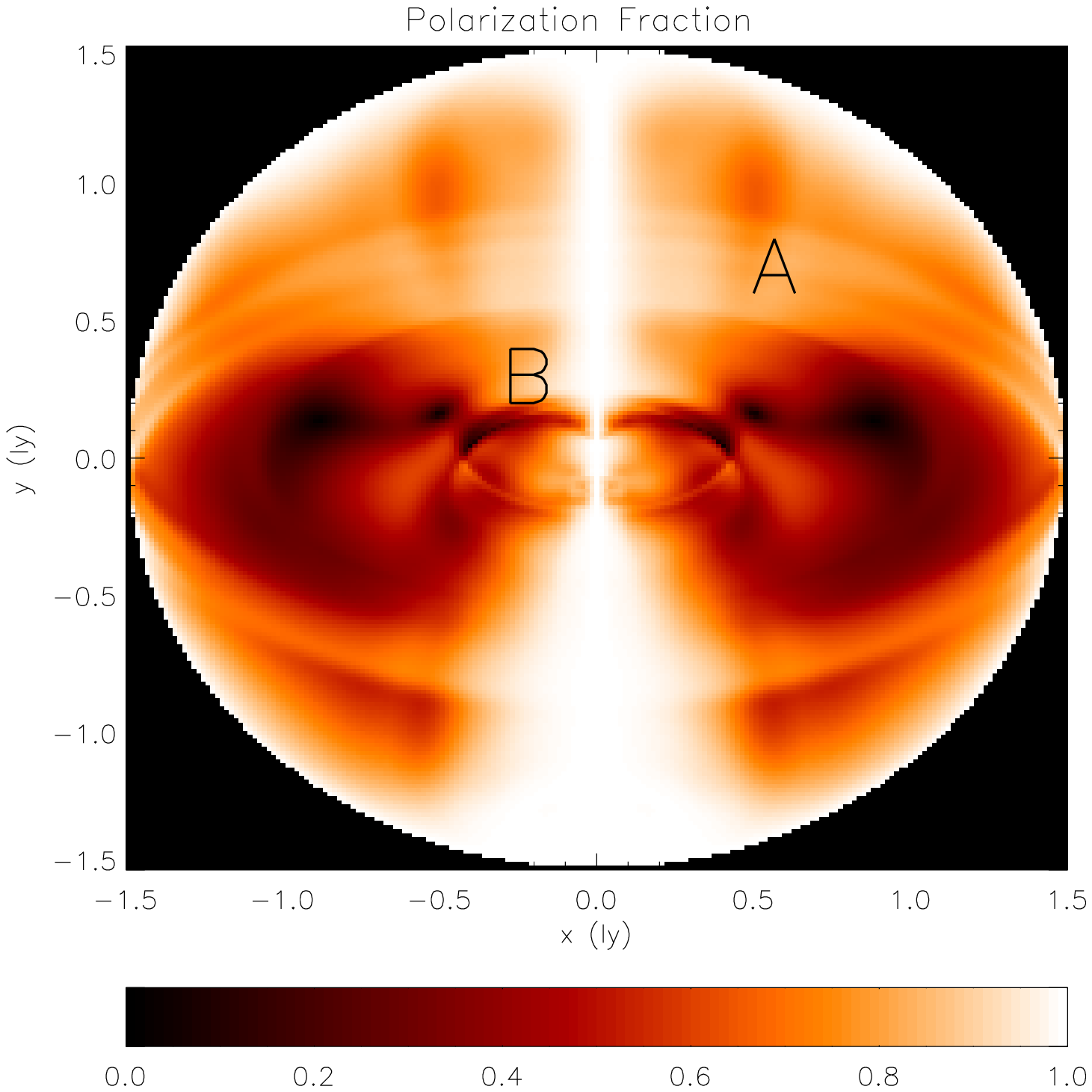}\includegraphics{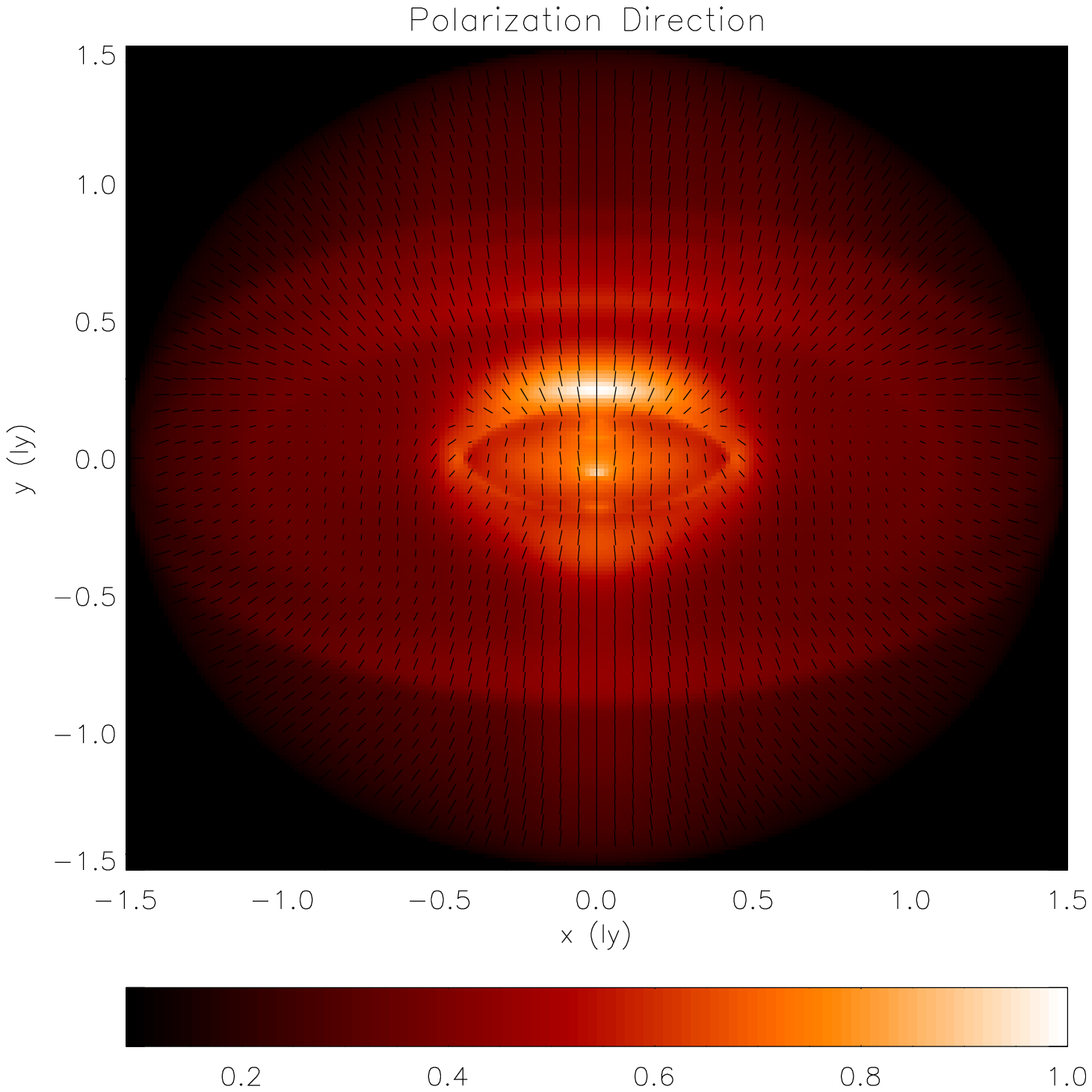}}
   \caption{Results from full relativistic MHD simulations. Left panel: polarization fraction in the inner region of a PWN, normalized to the theoretical maximum ($\sim 70\%$ for a particle distribution index $\alpha=0.5$). A: main torus; B: inner ring. Right panel: total optical emission (normalized to the maximum) and polarization vectors. The length of the vector is proportional to the polarization fraction. This map accounts for synchrotron losses, using typical values of age and size for the Crab Nebula.}
   \label{fig:2}
\end{figure*}
 
Concerning the polarization fraction we find that in the entire torus it is very close to its theoretical maximum. When the bulk flow exceeds $0.2\,c$, Doppler boosting effects are important and even where contributions from the front and back side of the torus are integrated together along the line of sight the depolarization is still negligible there. Minor deviations form the maximum polarization fraction are visible just on the outer sides. 

More interesting is the behavior of the polarization vector. As expected, close to the symmetry axis of the torus, coincident with that of the nebula, this is aligned with the axis itself (due to the assumption of a purely toroidal field), and the inclination of the polarization vector with respect to the symmetry axis increases moving away from it. The position where the polarization vector is perpendicular to the axis is a function of both the inclination of the torus with respect to the plane of the sky $\delta$, and the flow speed. 

We consider specifically an inclination of $\delta=30^\circ$, corresponding approximately to the best estimate for the Crab Nebula. We find that for increasing values of the bulk speed of 0.2, 0.4, 0.6, 0.8~$c$, the corresponding position at which the polarization vector is perpendicular to the axis moves closer in (see fig.~\ref{fig:1}). We estimate the position of this transition in a polar coordinate system centered on the pulsar and lying on the plane of the sky, with angles measured from the axis of the nebula. For the above mentioned values of velocity, the angle of the transition is $90^\circ$, $80^\circ$, $70^\circ$, $60^\circ$, respectively. 

Looking at Fig.~5 of Schmidt et al. (\cite{schmidt79}) we can notice in the regions labeled ``W1'' and ``W2'' that the electric field is perpendicular to the nebular axis at an angle of about $\sim 70^\circ$. 

 Besides, the polarization vector in the back side of the torus remains aligned with the axis, while diverges from it in the front side. This effect increases with bulk flow speed. Again, such difference in the divergence of the polarization vector between the front and back side of the nebula is observed in Fig.~5 of Schmidt et al. \cite{schmidt79}. If one wants to reproduce the observed front-back difference in the divergence of the electric field vector, a bulk velocity greater than at least $0.3\, c$ is required. We thus tentatively conclude that evidence for high flow velocities in the nebula was already present in that map.

\subsection{Numerical simulations}

Synthetic polarization maps to compare with the data can be obtained by applying the scheme outlined above to more sophisticated models for the magnetic field structure, flow velocity pattern, and emitting particle distribution. Axisymmetric relativistic MHD simulations (Del Zanna et al. \cite{luca04}) provide the full set of dynamical quantities. In order to properly evaluate the emission at high frequencies, as discussed in the previous section, what is left to take into account is the effect of synchrotron losses. In order to do this we have added to our code a tracer that evolves $\epsilon_\infty$ (see Eq.~(\ref{losses})) from the termination shock to the outer regions of the nebula following the streamlines, thus allowing us to model the modifications in the emitting particles' distribution function. 

We show in Fig.~\ref{fig:2} the results for the total polarization fraction and the polarization vector direction, at optical frequencies. Values are computed based on a recent simulation performed using a fixed spherical boundary condition (imposing a constant, non relativistic, flow speed $V_{\rm PWN}$ at the nebular outer boundary), thus neglecting the interaction of the PWN with the expanding ejecta. In the simulation we have used the following wind parameters: anisotropy $\alpha=0.1$, magnetization $\sigma=0.03$, and $b=10$ ($b$ parameterizes the width of the striped wind region of reduced magnetic field around the equator, see Del Zanna et al. \cite{luca04} for a precise definition of these parameters). The other quantities are typical of the Crab Nebula: luminosity $L=5\times 10^{38}$ erg~s$^{-1}$, age $t=1000$~yr, boundary speed $V_{\rm PWN}=1500$~km~s$^{-1}$. 

Let us discuss the results in Fig.~\ref{fig:2} and their implications. First of all we identify two main structures in the optical surface brightness map (right panel): an {\it inner ring} and an outer torus. As in the uniform ring case, the polarization reaches its maximum on the axis. The front side of the inner ring has a very high degree of polarization in a large region, up to about $60^\circ$ from the symmetry axis of the nebula, while, in the back side, the polarization tends to decline faster moving away from the axis. We also observe an enhancement of polarization in the region corresponding to the torus where values are of order 70\% of the maximum. Contrary to the toy-model case, where we found the entire torus to be highly polarized, now the polarization fraction drops below 50\% along the sides of the {\it inner ring}, and there are regions at about $80^\circ$ where the emission is almost depolarized. 

Concerning the direction of the polarization vector, we again observe the effect of Doppler boosting in the inner ring, where the polarization vector becomes perpendicular to the axis of the nebula at about $70^\circ$. Taking as reference our toy-model, this would imply a speed $\sim ~ 0.6 \, c$. The same effect is still present but less strong in the torus, where the bulk flow speed is much lower. It also seems that in the back side of the ring the polarization tends to remain parallel to the axis up to larger angular distances than in the front side.

In the outer region of the nebula the polarization vector is in the radial direction, as expected for a slow moving flow embedded in a toroidal field. However, the results in this region should not be taken too seriously because here our neglect of the interaction with the SNR material may be crucial. 

Our results suggest that the behavior of the polarization vector in the inner ring (i.e. the swing in position angle) might be used to constrain the flow speed, while the polarization fraction may help estimating the amount of disordered or poloidal magnetic field.

\section{Conclusions}
In this paper we have presented preliminary results concerning the properties of non-thermal emission of PWNe in 2-D also computing, for the first time the (linear) polarization structure. The latter task, on which this paper is mainly focused, has been carried out both in the framework of a simplified model and by using the results of relativistic MHD simulations of the flow inside the PWN (Del Zanna et al. \cite{luca04}). Our results suggest that the bulk flow in the emitting region and the polarization properties are strongly correlated. 

Our simplified model consists of a homogeneous torus with prescribed shape and inclination with respect to the plane of the sky. The torus has both uniform flow velocity, assumed to be purely radial, and magnetic field, purely toroidal. Under these conditions, we find that the polarization fraction is always close to the expected maximum, regardless of the radial speed, whereas, concerning the position angle, we have shown that for high velocities, due to relativistic effects, the polarization vectors diverge rapidly from the axis of the nebula, in the front side, but tend to remain aligned in the back side. 

A more complex pattern is found in the polarization maps based on numerical simulations. The synchrotron surface brightness map at optical frequencies shows two ring-like structures: a bright {\it inner ring}, at about the distance of the wind termination shock, and a fainter {\it torus} farther away from the pulsar. First of all we find a correlation between the intensity and the degree of polarization. This is however generally lower than in the previous case, with the emission almost completely depolarized at the edges of these brighter structures. As far as position angle is concerned, polarization swing is stronger in the inner ring than in the torus, due to a higher flow speed in the former, while both retain the front/back asymmetry found in the toy-model.

We would like to point out that this asymmetry was already showing in the observations of the Crab Nebula, and could have been used to infer relativistic speeds in the inner parts of the nebula. It should be clear at this point that the combination of optical  (and hopefully X-ray) high-resolution polarimetry, and synthetic polarization maps from relativistic MHD simulations, is promising to provide a powerful diagnostic tool for the flow structure in PWNe. Polarization measures might also help to clarify the origin of some puzzling emission features like the knot of the Crab Nebula, revealing whether they are associated with the nebular post-shock flow or a result of some kind of plasma instability in the pulsar wind. 

In order to understand how emission features and polarization properties can be used to infer the wind structure and magnetization, a more detailed study based on a set of numerical simulations is currently under progress. Results will be presented in a future paper.

\begin{acknowledgements}
This work has been partially carried out while three of us (N. B., L. D.Z., E. A.) were participating to the program {\it Physics of Astrophysical Outflows and Accretion Disks} hosted by the Kavli Institute for Theoretical Physics, of the University of California at Santa Barbara. We would like to thank KITP for hospitality and the scientific organizers of the program (O. Blaes, C. Gammie, H. Spruit). This research was supported in part by the national Science Foundation under Grant No.PHY99-07949. N.~B. was supported in part by NASA grant TM4-5000X to the University of California, Berkeley, and a David and Lucile Packward Fellowship to Eliot Quataert. 
\end{acknowledgements}


\end{document}